\documentclass[conference]{IEEEtran}
\IEEEoverridecommandlockouts
\usepackage{cite}
\usepackage{amsmath,amssymb,amsfonts}
\usepackage[ruled]{algorithm2e}
\usepackage{graphicx}
\usepackage{subfig}
\usepackage{textcomp}
\usepackage{multirow}
\usepackage{multicol}
\usepackage{arydshln}
\usepackage{float}
\usepackage{booktabs}
\usepackage[table,xcdraw]{xcolor}
\usepackage{array}
\usepackage{verbatim}

\def\BibTeX{{\rm B\kern-.05em{\sc i\kern-.025em b}\kern-.08em
    T\kern-.1667em\lower.7ex\hbox{E}\kern-.125emX}}
\begin{document}

\title{Research on Mobile Network High-precision Absolute Time Synchronization based on TAP}
\author{Chenyu~Zhang,~Xiangming~Wen,~Wei~Zheng,~Longdan~Yu,~Zhaoming~Lu,~and~Zhengying~Wang~
\thanks{This work was supported in part by the National Key R\&D Program of China under Project 2019YFB1803303, and in part by BUPT Excellent Ph.D. Students Foundation under Project CX2021212.}
\thanks{The authors are with the Beijing Laboratory of Advanced Information Networks and Beijing Key Lab of Network System Architecture and Convergence, Beijing University of Posts and Telecommunications, Beijing 100876,
China (e-mail: octopus@bupt.edu.cn, zhengweius@bupt.edu.cn).}
}

\maketitle
\begin{abstract}
With the development of mobile communication and industrial internet technologies, the demand for robust absolute time synchronization based on network for diverse scenarios is significantly growing. TAP is a novel network timing method that aims to achieve sub-microsecond synchronization over air interface. This paper investigates the improvement and end-to-end realization of TAP. This paper first analyzes the effectiveness and deficiencies of TAP by establishing an equivalent clock model which evaluates TAP from timing error composition and allan variance. Second, this paper proposes a detailed base station and terminal design and the corresponding improvement of TAP. Both hardware compensation and protocol software design are taken into account so as to minimize timing error and system cost while maximizing compatibility with 3GPP. Finally, this paper presents a TAP end-to-end 5G prototype system developed based on software defined radio base station and COTS baseband module. The field test results show that the proposed scheme effectively solves the problems of TAP in application and robustly achieves 200ns level timing accuracy in various situations. The average accuracy with long observations can reach 1 nanosecond. It is 2$\sim$3 orders of magnitude better than common network timing methods, including NTP, PTP and the original TAP.
\end{abstract}


\section{Introduction}

\IEEEPARstart{W}{ith} the development of wireless communication technology, the 5G mobile network has the characteristics of ultra-large bandwidth, ultra-low latency and massive access. Thus traditional and emerging applications in many vertical industries can be accessed through mobile networks\cite{a0.1}. For considerable applications in the fields of industry, internet of things, transportation, etc., absolute time synchronization is a basic requirement. The mobile network itself and the vertical applications have been putting forward higher and higher requirements for the accuracy of absolute time synchronization. For example, the further integration of mobile network and time-sensitive communication technology to realize wireless industrial control and monitoring is one of the hottest topics of mobile communication. Sub-microsecond time synchronization is essential for current technologies such as Time Sensitive-Networking (TSN). The same requirement will surely apply to the mobile network\cite{re1-1}. Some RF-based localization methods require nanosecond timing to complete the ranging or reflective mapping\cite{re1-2}. Some AI-enabled sensoring applications can use sub-microsecond timing to improve their efficiency\cite{re1-3}.
Aiming at these demands, the realization of robust time synchronization, time maintenance and time transfer in each wired and wireless node of the mobile network is the trend of B5G\cite{5GA} and 6G network\cite{a0}\cite{a1}\cite{add1}. 3GPP has proposed mechanisms in Release 16 where absolute timing is a major improving issue\cite{a1.1}.

\begin{table*}[t!]
  \centering
  \caption{Comparison of the existing timing methods from the practical engineering realization perspective.}
  \label{table1}
  \begin{tabular}{|cc|ccc|c|c|c|ccc|}
    \hline
    \multicolumn{2}{|c|}{\multirow{2}{*}{}} &
      \multicolumn{2}{c|}{TAP} &
      \multirow{2}{*}{TAPv2} &
      \multirow{2}{*}{NTP} &
      \multirow{2}{*}{PTP} &
      \multirow{2}{*}{GNSS} &
      \multicolumn{1}{c|}{\multirow{2}{*}{FTSP}} &
      \multicolumn{1}{c|}{\multirow{2}{*}{RBIS}} &
      \multirow{2}{*}{WizSync} \\ \cline{3-4}
    \multicolumn{2}{|c|}{} &
      \multicolumn{1}{c|}{Original} &
      \multicolumn{1}{c|}{This paper} &
       &
       &
       &
       &
      \multicolumn{1}{c|}{} &
      \multicolumn{1}{c|}{} &
       \\ \hline
    \multicolumn{2}{|c|}{\begin{tabular}[c]{@{}c@{}}Typical\\ Scenarios\end{tabular}} &
      \multicolumn{3}{c|}{\begin{tabular}[c]{@{}c@{}}Mobile network\\ \textless{}10km\end{tabular}} &
      Layer 3 &
      Layer 2 &
      \begin{tabular}[c]{@{}c@{}}Satellite\\covered area\end{tabular} &
      \multicolumn{1}{c|}{\begin{tabular}[c]{@{}c@{}}ZigBee\\ \textless{}0.3km\end{tabular}} &
      \multicolumn{2}{c|}{\begin{tabular}[c]{@{}c@{}}WiFi\\ \textless{}0.3km\end{tabular}} \\ \hline
    \multicolumn{1}{|c|}{\multirow{2}{*}{Accuracy}} &
      typical &
      \multicolumn{1}{c|}{\multirow{2}{*}{\begin{tabular}[c]{@{}c@{}}$10^{-6}$\\ $\sim$$10^{-5}s$\end{tabular}}} &
      \multicolumn{1}{c|}{\multirow{2}{*}{\begin{tabular}[c]{@{}c@{}}$10^{-8}$\\ $\sim$$10^{-6}$s\end{tabular}}} &
      \multirow{2}{*}{\begin{tabular}[c]{@{}c@{}}$10^{-9}$\\ $\sim$$10^{-6}$s\end{tabular}} &
      \begin{tabular}[c]{@{}c@{}}$10^{-4}$\\ $\sim$$10^{-1}$s\end{tabular} &
      \begin{tabular}[c]{@{}c@{}}$10^{-9}$\\ $\sim$$10^{-7}$s\end{tabular} &
      \multirow{2}{*}{\begin{tabular}[c]{@{}c@{}}$10^{-10}$\\ $\sim$$10^{-7}$s\end{tabular}} &
      \multicolumn{1}{c|}{\begin{tabular}[c]{@{}c@{}}$10^{-7}$\\ $\sim$$10^{-6}$s\end{tabular}} &
      \multicolumn{1}{c|}{\begin{tabular}[c]{@{}c@{}}$10^{-6}$\\ $\sim$$10^{-5}$s\end{tabular}} &
      \begin{tabular}[c]{@{}c@{}}$10^{-5}$\\ $\sim$$10^{-4}$s\end{tabular} \\ \cline{2-2} \cline{6-7} \cline{9-11} 
    \multicolumn{1}{|c|}{} &
      \begin{tabular}[c]{@{}c@{}}mobile\\network\end{tabular} &
      \multicolumn{1}{c|}{} &
      \multicolumn{1}{c|}{} &
       &
      \begin{tabular}[c]{@{}c@{}}$10^{-2}$\\ $\sim$$10^{-1}$s\end{tabular} &
      \begin{tabular}[c]{@{}c@{}}$10^{-4}$\\ $\sim$$10^{-2}$s\end{tabular} &
       &
      \multicolumn{3}{c|}{incompatible} \\ \hline

    \multicolumn{2}{|c|}{\begin{tabular}[c]{@{}c@{}}Typical\\ observation time\end{tabular}} &
    \multicolumn{1}{c|}{10s} &
    \multicolumn{1}{c|}{2.5s} &
    1s &
    1s &
    5s &
    1min$\sim$hours &
    \multicolumn{1}{c|}{\textless{}1s} &
    \multicolumn{1}{c|}{\textless{}1s} &
    \textless{}1s \\ \hline

    \multicolumn{2}{|c|}{\begin{tabular}[c]{@{}c@{}}Compatible with \\ 3GPP R16\end{tabular}} &
      \multicolumn{2}{c|}{Y} &
      X &
      Y &
      Y &
      / &
      \multicolumn{1}{c|}{/} &
      \multicolumn{1}{c|}{/} &
      / \\ \hline
    \multicolumn{2}{|c|}{\begin{tabular}[c]{@{}c@{}}Frame offset\\measurement/config\end{tabular}} &
      \multicolumn{2}{c|}{required} &
      optional&
      \begin{tabular}[c]{@{}c@{}}not\\required\end{tabular} &
      \begin{tabular}[c]{@{}c@{}}not\\required\end{tabular} &
      / &
      \multicolumn{1}{c|}{\begin{tabular}[c]{@{}c@{}}not\\required\end{tabular}} &
      \multicolumn{1}{c|}{\begin{tabular}[c]{@{}c@{}}not\\required\end{tabular}} &
      required\\ \hline
    \multicolumn{2}{|c|}{\begin{tabular}[c]{@{}c@{}}Optimized \\for mobility\end{tabular}} &
      \multicolumn{2}{c|}{low-speed} &
      high-speed &
      X &
      X &
      high-speed &
      \multicolumn{1}{c|}{X} &
      \multicolumn{1}{c|}{X} &
      X \\ \hline
    \end{tabular}
  \end{table*}

Absolute time synchronization technology has been studied for decades, such as the ubiquitous Network Time Protocol (NTP)\cite{a2}, the widely-used Precision Timing Protocol (PTP)\cite{re1-4}, the Global Navigation Satellite System (GNSS) timing, the long and short wave timing (like low-frequency-time-code), the Flooding Time Synchronization Protocol (FTSP)\cite{re1-5} for Zigbee radios, Reference Broadcast Infrastructure Synchronization (RBIS)\cite{re1-6}, WizSync for Wireless Sensor Networks (WSN)\cite{re1-7}, etc\cite{a3}\cite{a3.1}. These technologies and their derivatives can well meet the needs of traditional application\cite{a4}\cite{a4.0}. 
Specifically, the GNSS-based timing methods are suitable for outdoor nanosecond level timing. The method proposed in \cite{a4.3} is used in satellite-ground time-frequency communication, achieving upto 0.5 picosecond ($10^{-12}s$) synchronization. These methods require an open area for good satellite signal reception. 
By contrast, the network-based timing performs much better in terms of security, scenario diversity and cost control. The method proposed in \cite{a4.1} can achieve $10^{-4}s$ synchronization with short observation in wireless network by applying matrix completion-based maximum likelihood estimation. It is designed for low-cost and energy-sensitive WSN devices. The method proposed in \cite{a4.2} can achieve $10^{-5}s$ synchronization by applying bayesian estimation. The mechanism proposed in \cite{re1-7.1} discussed about the possibility of sub-microsecond timing based on LTE system in close range. The FTSP, RBIS and WizSync can achieve sub-microsecond accuracy in hundreds of meters of communication. However, these broadcast timing methods do not compensate for transmission delay. They are not suitable for mobile network which has much longer communication distance.

Different from other wireless networks, mobile network timing has the following characteristics:
\begin{enumerate}
  \item The longer communication distance requires robust delay compensation;
  \item The air interface protocol stack reduces the delay compensation accuracy of technologies such as PTP;
  \item The larger and complex topologies requires low hardware and bandwidth overhead of timing functions;
  \item The high-precision absolute time delivery should basically be compatible with 3GPP standards.
\end{enumerate}
Therefore, achieving sub-microsecond timing accuracy in mobile network with large-scale equipments in diverse scenarios at low cost challenges all the traditional methods\cite{a4.01}\cite{hx1}. Their inability to simultaneously meet the demands in terms of precision, flexibility and cost has greatly hindered the realization of the vertical applications mentioned above.

The High-precision Timing Method over Air Interface based on Physical-Layer Signals (TAP) is a novel mobile network timing method that aims to achieve sub-microsecond time synchronization over the standard air interface\cite{tap}. It reuses mobile network software \& hardware and applies physical layer signals of air interface to directly transmit time information, so as to achieve high-precision timing while keeping costs relatively low. Similar to TAP, some methods based on mobile networks were discussed as well\cite{taplike1}\cite{taplike2}. However, those works of TAP (and its improved schemes) only discuss the feasibility in terms of simulation/theory\cite{tapv2}. The technical indicators and applicable scenarios of the mentioned methods are shown in Table \ref{table1}. 

As network timing is a highly engineering and application-oriented technology, like NTP and PTP, its effectiveness can only be verified through an end-to-end(E2E) system. The purpose of this paper is to discuss how to implement the TAP E2E system with sub-microsecond accuracy and robustness for 5G network. The main contributions of this paper are as follows:
\begin{itemize}
    \item We propose an equivalent TAP-clock model and a mobile network timing error model to analyse the end to end realization of TAP. We theoretically discuss the inadequacies of the original TAP and improve the design of its mechanism and statistical compensation for clock source error, time information transmitting error and receiving error. We accordingly propose an Allan variance optimization method for TAP. 
    \item We propose a detailed Base Station (BS) and User Equipment (UE) scheme focusing on ensuring sub-microsecond timing accuracy, maximizing compatibility with 3GPP standards and minimizing system costs. We discuss the timing and compensation process, UE calibration and protocol software design from an engineering point of view corresponding to the clock model. 
    \item We present a prototype system of TAP for the first time based on software defined radio and Commercial Off-The-Shelf (COTS) UEs. The field test verifies the feasibility of low-cost and high-precision mobile timing, as well as the effectiveness of the proposed schemes.
\end{itemize}

The article is structured as follows. In Section II, we describe the principle of PTP and TAP. In Section III, we describe the mobile network timing error model together with the related algorithms. In Section IV, we discuss the design of TAP implementation. In Section V, we carry out some field test of the implemented prototype system and analyze its performance. Finally we conclude this paper in Section VI.

\begin{figure}[t!]
  \begin{center}
      \includegraphics[width=1.5in]{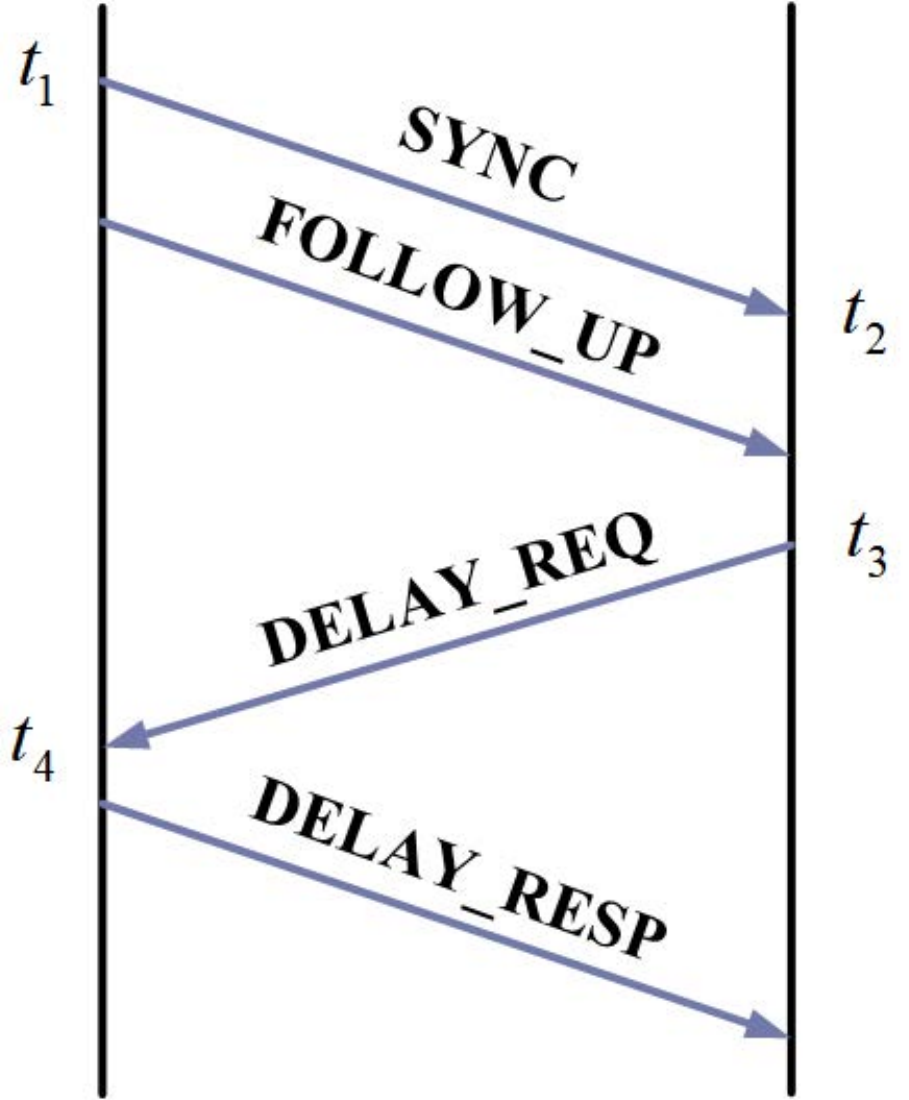}
  \end{center}
  \caption{The delay compensation of PTP.}
  \label{fig-PTP}
\end{figure}

\section{Related Work}
\subsection{Precision Time Protocol}
PTP is a representative high-precision network timing method with delay compensation. It is based on Ethernet's physical layer timestamp and uses the mechanism shown in Fig. \ref{fig-PTP} to compensate for the delay\cite{re1-4}. PTP compensates the downlink delay through the sending and receiving timestamps of SYNC and DELAY\_REQ messages. The offset between the master and the slave clock is:
\begin{equation}
  t_{offset}=t_2-[t_1+(t_2-t_1+t_4-t_3)/2]
\end{equation}
Together with hardware and statistical compensation algorithms like Kalman filtering\cite{a3} and White Rabbit\cite{re1-8}, PTP is presently the most accurate E2E timing method in wired network, especially in optical networks. Many wireless timing methods are optimized for specific networks, applying PTP in those networks is often used as a benchmark\cite{re1-9}. As PTP relies on the symmetry of Ethernet, it is usually several orders of magnitude less accurate in wireless networks.

\subsection{TAP}
TAP was first proposed in \cite{tap} and improved in \cite{tapv2} as TAPv2. This section describes their principle and discusses about their insufficiency for E2E timing.
\subsubsection{Scenario}
The mobile network timing has the following advantages comparing with other timing methods:
\begin{itemize}
  \item It is easy to deploy a high-precision master clock in mobile network to synchronize the nodes.
  \item The forms of base station are diverse, such as macro BS, micro BS, pico BS and indoor distribution systems. The signal can be well covered in various scenarios.
  \item The air interface has a relative synchronization mechanism, which can provide a lower limit guarantee for timing accuracy.
\end{itemize}

Thus the applicable scenarios of TAP mainly include:
\begin{itemize}
  \item Areas without GNSS signal coverage, such as indoor, underground, mine, tunnel. 
  \item Terminals without GNSS module.
  \item Low-cost wireless terminals which require timing accuracy better than 1 microsecond, such as industrial IoT sensors and wireless time-sensitive communication terminals.
\end{itemize}

\subsubsection{Method}
\begin{figure}[t!]
  \begin{center}
      \includegraphics[width=3.5in]{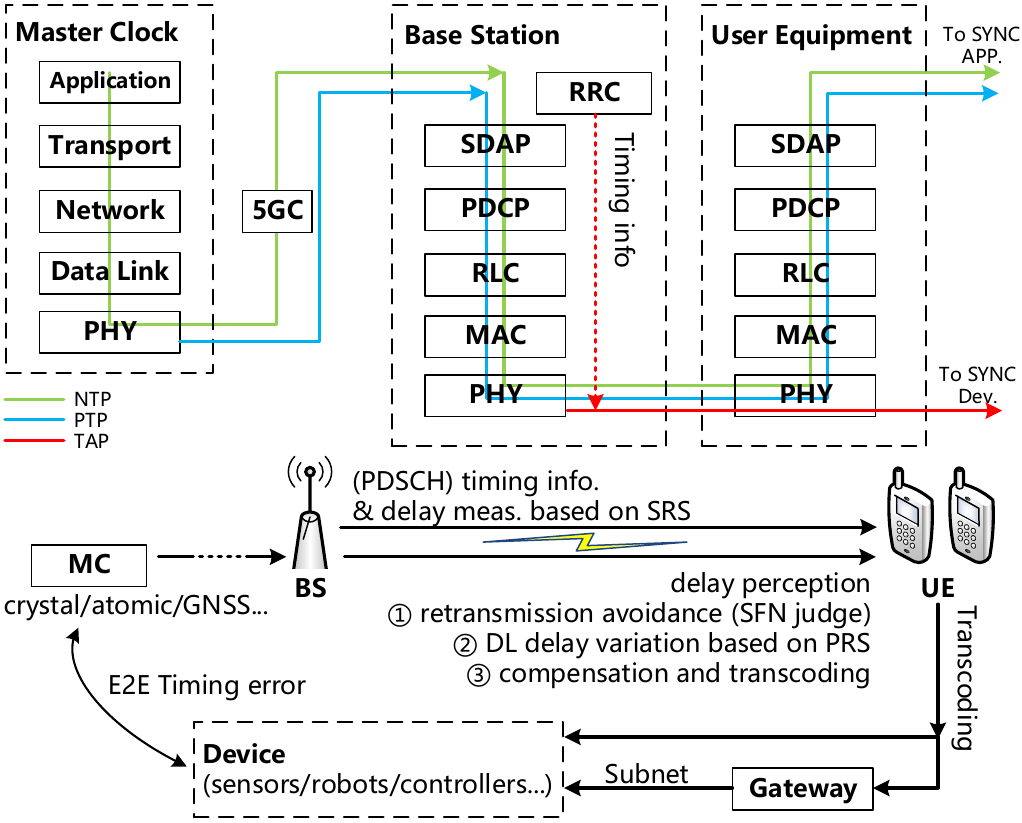}
  \end{center}
  \caption{The principle and scenario of TAP compared with NTP and PTP.}
  \label{fig-tap}
\end{figure}

TAP follows the classic timing principle, i.e. accurately sending the timestamp and compensating the transmission time through delay estimate which is usually based on the link symmetry. TAP takes advantage of physical layer signals of air interface, which gives it the ability of accurately estimating downlink delay. As shown in Fig.\ref{fig-tap}, to avoid the asymmetry of the processing delay of the wireless protocol stack, its timing process is carried by the PHY layer. It applies downlink signaling like MIB or SIB to carry the absolute time information. It applies timing advance (TA) for delay estimation. And the time phase synchronization is based on the measurement of relative synchronization of the air interface. It applies a two-threshold-control for a smooth timing convergence. The timing result will be transcoded for the hardware clock or downstream equipments. The simplified formula for calculating the timing result when applying TA for delay estimation can be expressed as:
\begin{equation}
  \begin{aligned}
    T_t = T_{BS} + t_{est} + t_0 \\
    t_{offset} = T_{ue} - T_t      
  \end{aligned}
\end{equation}
where $T_{BS}$ represents the BS time. $t_{est}$ represents the estimated compensation of channel delay, it equals to $TA/2$ here. $t_0$ represents the processing delay of time information (from baseband to output), which is essentially a part of the timing asymmetry that can be measured and eliminated. $T_t$ is the absolute timing result which can be used for transcoding and compensation algorithm. $t_{offset}$ is the difference between timing result and the UE local time $T_{ue}$. The timing error is mainly caused by the randomness of wireless channels, which directly affects the estimation of $TA$. Generally, TAP provides a basic mechanism for high-precision timing in mobile networks. 

\subsection{TAPv2}
TAPv2 is an improved version of TAP based on 3GPP R16 standard. It mainly contains improvements from three aspects.
First, it applies System Information Block 9 (SIB9) defined in TS38.331\cite{38331} to carry timing info. The timestamp generated by RRC layer is not ``stamped" but predicted according to the MAC scheduling results. The granularity of absolute time is extended to $10ns$, and the reference system frame number (SFN) is used to eliminate the error brought by retransmission. 

Second, as the accuracy of the delay estimation is determined by the pilot signal, it applies Sounding Reference Signal (SRS) and Positioning Reference Signal (PRS) signal for delay perception instead of TA. As defined in 3GPP TS38.21x\cite{38213}, SRS is generated by cyclic shifts of ZC sequences with good correlation characteristics:
\begin{equation}
  ZC_u(n) = e^{-j\pi un(n+1)/N}
\end{equation}
where $u$ is the root index of the sequence and $N$ is the length of the sequence. As also defined in 38.21x, PRS is generated by modulating a binary pseudo-random sequence with good autocorrelation and cross-correlation characteristics:
\begin{equation}
  \begin{aligned}
    S(n) = \frac{1}{\sqrt{2}}(1-2c(2m))+j\frac{1}{\sqrt{2}}(1-2c(2m+1)) \\m=0,1,...,2N-1
  \end{aligned}
\end{equation}
where $c(m)$ is a pseudo-random sequence and $N$ is the maximum number of resource blocks in the system link. Thus, the result of delay estimation is no longer in units of $16T_s$, but can be accurate to the $T_c$ level, where $T_c$ is the minimum time interval for the 5G system and $T_s$ is several times of $T_c$ according to the numerology configuration.

Third, to improve the applicability of the algorithm, it proposes a precise estimation algorithm of SRS and PRS to improve the accuracy and robustness of the estimation for the high-speed terminals. And a deep reinforcement learning based method is applied to optimize the settings of parameters in TAP, so as to automatically adapt to various channel environments.

In our opinion, TAP provides an effective mobile network timing paradigm due to the timing demands of vertical applications on mobility, low cost and high precision. The idea of applying air interface PHY signals for timing and the design of timing signaling is valuable. Based on these works, there are still the following issues to be resolved before TAP can be put into practical use: 1) The evaluation model and theoretical analysis of TAP, so as to extend it to E2E system; 2) Timing hardware and software design, compatible with 3GPP standards and ensure robustness timing accuracy; 3) Evaluation and tests based on prototype system and actual Radio Frequency (RF). These issues are exactly what this paper intends to address.

\begin{figure}[t!]
  \begin{center}
      \includegraphics[width=3.5in]{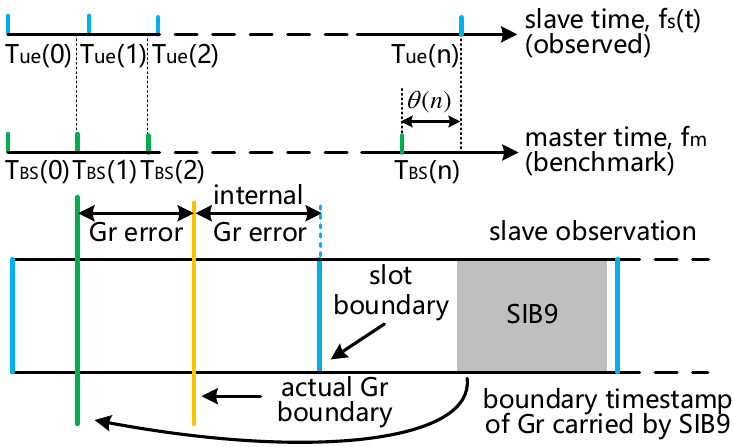}
  \end{center}
  \caption{The proposed equivalent TAP-clock model.}
  \label{fig-tc}
\end{figure}

\section{E2E Analysis of TAP Realization}
In this section, we discuss the rationale for precision optimization in TAP realization. From the perspective of observation, we model TAP as a composite clock model which is feasible of applying Allan variance optimization. From the perspective of time transfer, we propose a mobile network timing error model to provide a ``segmented" optimization route for the E2E implementation of TAP.

\subsection{Equivalent TAP-clock Model}
Generally speaking, Master Clock (MC) must have robust timekeeping capabilities, such as high-stability crystal oscillators, atomic clocks, GNSS (atomic clocks carried by satellite signals). But the base stations in TAP timing play the role of Slave Clock (SC) or transparent clock, it relies on other clock sources and introduces random errors due to the air interface. Therefore, we propose an equivalent clock model to describe TAP.

The network clock source is triggered by discrete ``events'', such as packet arrivals, symbol boundaries, changes in frequency and phase are mixed while a certain degree of uncertainty that related to the system sampling rate affects these events. For TAP, the corresponding ``event'' is sending/receiving SIB9 signaling. As shown in Fig.\ref{fig-tc}, assuming that the event triggered by BS is at $T_{BS}(t)$ and the event observed by UE is at $T_{ue}(t)$, we can express the UE-observed TAP-clock as: 
\begin{equation}
  \begin{aligned}
    df_s|_t = df_{src} + df_{ue} + n_{\phi}(t) + n_{s} \\
    dT_{ue}|_{t}/f_s(t) = dT_{BS}|_t / f_m
    \label{f5}
  \end{aligned}
\end{equation}
where $f_s$ represents the observed frequency, it is influenced by the uncertainty of BS's clock $df_{src}$, the uncertainty of UE's clock itself $df_{ue}$, sampling ambiguity $n_{s}$ and the interference caused by air interface sending and receiving $n_{\phi}(t)=n_{\phi}^{tx}(t)+n_{\phi}^{rx}(t)$. The frequency of BS clock can be regarded as a constant $f_m$ as it is the reference of timing. Let the cumulative deviation between UE and BS clock at time $t$ be $\theta(t)$, then we can obtain the following formula:

\begin{equation}
  \begin{aligned}
    n_s+\int_{0}^{T_{BS}(t)} f_s(t)\, dT_{ue} + n[T_{BS}(t)] \\
    = f_m{\int_{0}^{T_{BS}(t)}\, dT_{BS} +\theta[T_{BS}(t)]} \\
  \end{aligned}
  \label{f6}
\end{equation}
where $n[T_{BS}(t)]$ is the measurement noise of $T_{ue}(t)$ caused by UE's oscillator. The timing error can be represented by $\theta$, which is impossible for UE to obtaine. Only $d\theta$ can be calculated from each observation result:
\begin{equation}
    d\theta|_{T_{BS}(t)} = \frac{f_s[T_{BS}(t)]-f_m}{f_m}dT_{BS}|_{t} + \frac{1}{f_m}n_{\theta}[T_{BS}(t)]
    \label{f7}
\end{equation}
Then according to the definition of allan variance:
\begin{equation}
  \begin{aligned}
    \delta _{\theta}^2(T) = \frac{1}{2T}\mathbb{E} \big( \int^T_0\frac{d\theta}{dt}|_{T_{BS}(t)-t^{\prime}}dt^{\prime}
    - \int_0^T\frac{d\theta}{dt}|_{T_{BS}(t)-T-t^{\prime}}dt^{\prime} \big)^2
  \end{aligned}
  \label{f8}
\end{equation}
The UEs can obtain the change and allan variance of $\theta$ through the observed $t_{offset}$, thus evaluating the TAP clock.
$t_{offset}$ can be expressed as the following formula according to \cite{tapv2}: 
\begin{equation}
  \begin{aligned}
    t_{offset}(t) = T_{ue}(t) - T_{BS}(t) - t_{est}|_{abs[d_{SRS}/2-d_{PRS}]<\epsilon dt} - t_0
  \end{aligned}
  \label{f9}
\end{equation}
where $\epsilon$ stands for the control threshold of PRS-assisted timing error sensing. 

Applying formula (\ref{f7}) and (\ref{f8}), we can describe the timing quality of air interface, but what is its physical meaning? i.e. how does the equivalent clock relate to the actual engineering? Looking back into Fig.\ref{fig-tc}, the absolute timestamp carried by SIB9 indicates the boundary of the slot. It is assigned to a certain granularity unit $Gr$ which is defined as 10$ms$ in R15 and 10$ns$ in R16. Its accuracy is determined by the reference clock of the BS. The gap between the actual boundary and the timestamp of boundary is referred to $e_{Gr}$, which is apparently in unit of $Gr$. To reduce this error, it is necessary to jointly optimize the BS's clock, RRC timestamp generation and MAC scheduler. The gap between the actual boundary and the slot boundary is referred to $e_{inGr}$, which is introduced by the RF of the base station. Optimizing the clock trigger and fronthaul interface delay is expected to reduce its error. But generally the dedicated hardware-based PHY and RF are relatively stable, there is not much room for optimization. When $Gr$ is large enough, we can separate the $df_{src}$ that corresponding to $e_{Gr}$ and the $n_{\theta}^{tx}$ that corresponding to $e_{inGr}$. 

\begin{figure*}[t]
  \begin{center}
      \includegraphics[width=7.2in]{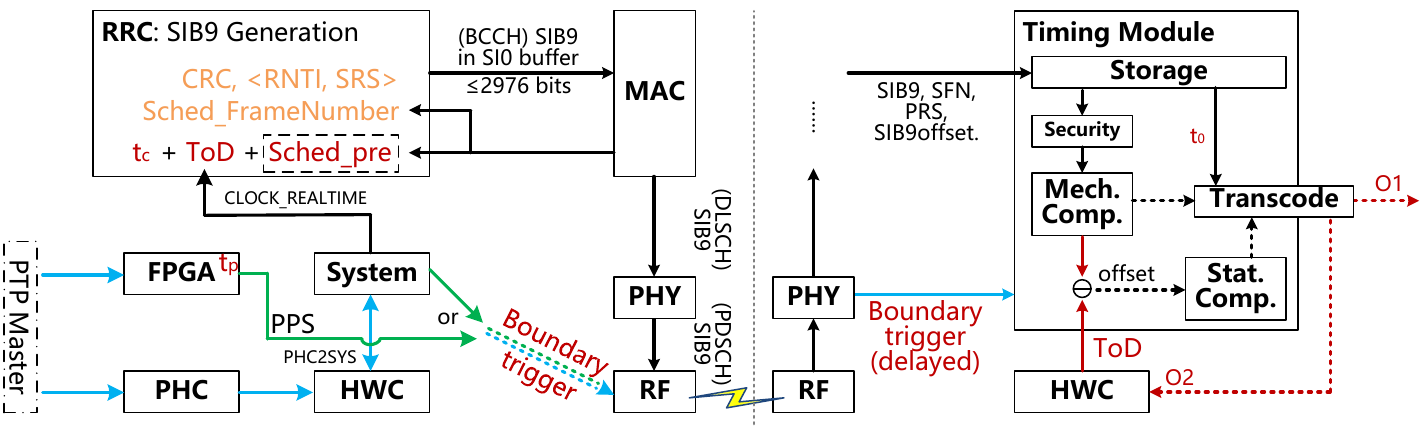}
  \end{center}
  \caption{Design of TAP implementation for BS and UEs.}
  \label{fig-bsue}
\end{figure*}
\subsection{Analysis of TAP Realization}
The mobile network E2E timing error can be expressed as:
\begin{equation}
  \begin{aligned}
    e = e_{src} + e_{node} + e_{Gr} + e_{inGr} + e_{air} + t_0
  \end{aligned} 
\end{equation}
\subsubsection{$e$} the difference between the calculated absolute time $t_{offset} + T_{ue}$ and the master clock. 
\subsubsection{$e_{src}$} Source error. At present, COTS rubidium clocks can be miniaturized and cost-effective\cite{b0}, which is suitable for MEC to enhance the time-maintenance capability at the mobile edge, reducing error and increasing stability.
\subsubsection{$e_{Gr}$ and $e_{inGr}$} Errors from the multi-layers of air interface protocol stack. A larger $Gr$ can reduce the complexity of the protocol software implementation, and it is easier to reduce $e_{Gr}$ to 0. A smaller $Gr$ has a better lower limit of timing accuracy and the system complexity is lower. But it has high requirements on the scheduler. The multiple timing strategies in \cite{tap} are equivalent to the selection of different $Gr$. We believe that the following should be considered for $Gr$: 1) compatibility with the minimum 5G standard; 2) simplifying the scheduling of SIB and reducing computational overhead; 3) avoiding hardware changes. Based on the characteristics of the BS's clock source, we believe that $10ms$ is currently the most suitable choice, as discussed in Section IV.
\subsubsection{$e_{air}$} TAP compensates for the error caused by air interface through multiple processes. We propose to divide the compensation into two parts: the mechanism compensation and the statistical compensation. The former determines the accuracy level, which includes the process of sending SIB and estimating downlink delay based on TA/SRS/PRS. The multi-state control and reference SFN judgment proposed in TAP are simple and effective, while the SRS/PRS precise estimation has high computational complexity and is not practical in most low-speed scenarios. Since SIB is broadcast signal, we propose to transmit the key-value pair of the UE's Radio Network Tempory Identity (RNTI) and feedback in the ``$nonCriticalExtension$'' field of SIB9. The statistical compensation improves robustness and the average accuracy. TAP sets a number of observations in a single timing process $K_0$ according to test experience, it takes the average accuracy as the optimization target. We suppose it is equivalent to the Allan variance optimization based on formula (\ref{f8}). Methods like Kalman filtering, Bayesian estimating, K-means, linear regression, etc. are also effective. The upper bound on air interface delay estimation is related to the bandwidth and sampling rate of the system. And the protocol guarantees its lower bound will not be worse than the length of the cycle prefix (CP). For the 5G system with $\mu=1$ (i.e. 30KHz subcarrier spacing and 4096-point FFT), the symbol sampling interval is $\Delta t=1/SCS/N_{FFT}\approx 8ns$ and $CP=2.34\mu s$. That is, The range of $e_{air}$ is [$\pm 4ns,\pm 2.34\mu s$].
\subsubsection{$t_0$} In the process of receiving timing information, all calculations and time format transcoding will introduce additional delay and uncertainty, leading to $t_0$. Taking outputting IRIG-B format for example, the YBC400X series chips that commonly used in low-cost devices has a delay of 1$\sim$3$\mu s$, which is already an order of magnitude failing to reach the timing accuracy target. We suppose that using offset for statistical compensation calculation, reducing the times of obtaining local timestamps and setting a fixed-step output delay can stabilize this delay. Then, $t_0$ can be compensated by the measurement and calibration UE at the factory.

We verify the plausibility of the proposed model and methods through field tests in Section V.

\begin{figure}[t!]
  \begin{center}
      \includegraphics[width=3in]{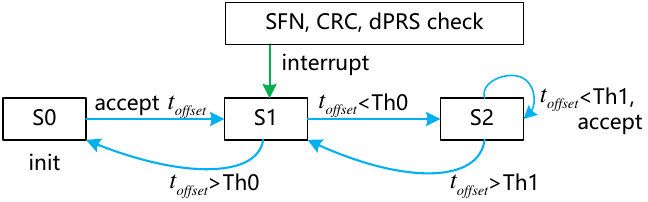}
  \end{center}
  \caption{Workflow of the state machine in timing module.}
  \label{fig-ue}
\end{figure}
\section{E2E Implementation of TAP}
In this section, we propose a detailed BS and UE design focusing on maximizing compatibility with 3GPP standards, minimizing system costs and ensuring sub-microsecond timing accuracy. Algorithm \ref{algo4-1} summarizes the workflow of the proposed methods.

\begin{algorithm}
    \DontPrintSemicolon

      BS synchronizes to reference clock.\;
      UE connects to the BS.\;
      \tcc{(Section IV.A)}
      BS generates SIB9: 1) BS predicts the absolute time of SIB9 according to the reference clock. 2) BS schedules SIB9. 3) BS calculates $Sched_pre$ and $Sched_FrameNumber$ accordingly. 4) BS calculates the $timeinfoUTC$ of SIB9. 5) BS fills the extension field of SIB9 according to the estimated downlink delay of UE. 6) BS calculates CRC of the entire signaling.\;
      
      BS triggers the RF according to reference clock and sends SIB9\;
      \tcc{(Section IV.B)}
      UE performs the initial calibration process to calculate $t_0$\;
      UE's baseband decodes SIB9 and send the SFN, TA, SIB9 offset to the timing module.\;
      UE performs mechanism compensation including reference-SFN check, PRS-based error sensing, CRC check and dual threshold control. (Fig. \ref{fig-ue})\;
      
      UE performs statistical compensation algorithms such as Allan variance optimizing and Kalman filtering.\;
      
      UE outputs the timing result, i.e. the absolute time.
      
    \caption{Workflow of the implemented TAP method.}
    \label{algo4-1}
\end{algorithm}

\subsection{BS: High-precision Time Delivery and Feedback}

Since absolute time synchronization is an optional function of BS, we design the TAP function as a plug-in service which is triggered by clock event. To reduce the error when the time information is sent from the BS, the designing steps include deciding granularity of timestamp, generating SIB9 and pre-compensating. 
The architecture diagram of BS is shown on the left in Fig. \ref{fig-bsue}, where PHC stands for PTP hardware clock, HWC stands for hardware clock of the BS, ToD stands for the timestamp truncated to $Gr$ when generating SIB9, $t_c$ is the difference between ToD and the slot boundary, Sched\_pre and Sched\_FrameNumber stands for the time delay and frame number of SIB9 that scheduled by MAC. 

First, the BS gets absolute time from reference clock source like PTP server. Both the hardware clock and system clock are synchronized to the reference. 

Second, the RRC application periodically generates timing signaling SIB9. We propose a redesigned SIB9 signaling compared with the standard. The \textit{timeinfoUTC} consists of the system timestamp ToD, the scheduling delay related to MAC Sched\_pre and $t_c$. The error of ToD, i.e. the error caused by software timestamping, is significantly worse than sub-microsecond. This error directly leads to $e_{Gr}$ while $t_c$ is used for its compensation. For the setting of $Gr$, we consider two cases. 
\begin{enumerate}
  \item $Gr=10ms$, which is compatible with R15: $e_{Gr}$ will not be introduced when generating SIB9, and $e_{inGr}$ is completely dependent on the accuracy of the trigger.
  \item $Gr=10ns$ which is compatible with R16: the RF can guarantee a small $e_{inGr}$ without relying on the trigger of the reference clock, but paradoxically the accuracy of $e_{Gr}$ depends on the reference clock.
\end{enumerate}
Considering that IoT devices are usually cost-sensitive, their basebands often only require minimum necessary functions, we suppose that setting $Gr$ to $10ms$ is the best choice.
For the extension field, the 32bit-RNTI-SRS pairs are filled in so as to provide precise delay estimation for UEs. (Terminals that do not require precise estimation can simply ignore the extension field.) The MAC interacts with the SIB generating module through an interface file with a mutex. In order to reduce the asymmetry caused by the delay of protocol stack, the baseband can process the received signaling without going through the upper layer CRC check. Thus an additional CRC of the generated signaling is also filled in the extension field so as to avoid timing fraud based on replay and selective forwarding mentioned in \cite{b1}. Every SIB9 can provide TA-based compensation for all UEs in the cell. And limited by the length restriction of SI which is 2976 bits, every SIB9 can provide SRS-based compensation for up to 88 UEs. i.e., each cell can serve 16500 times of high precision delay estimation per minute. 

Third, the reference clock generates PPS signal to trigger RF. In formula (\ref{f6}) we assume that $df_m\rightarrow 0$, i.e. the delay from trigger to the RF frame boundary is small enough. Commonly it is realized by GNSS-PPS trigger. But as there is no GNSS for the TAP scheme, we suppose a feasible solution is to apply an FPGA to convert PTP to PPS with a delay of $t_p$ so as to compensate for $e_{inGr}$. A detailed tutorial of building the proposed timing BS scheme can be found at \cite{opentap}.

\begin{figure*}[t!]
  \begin{center}
      \includegraphics[width=5.5in]{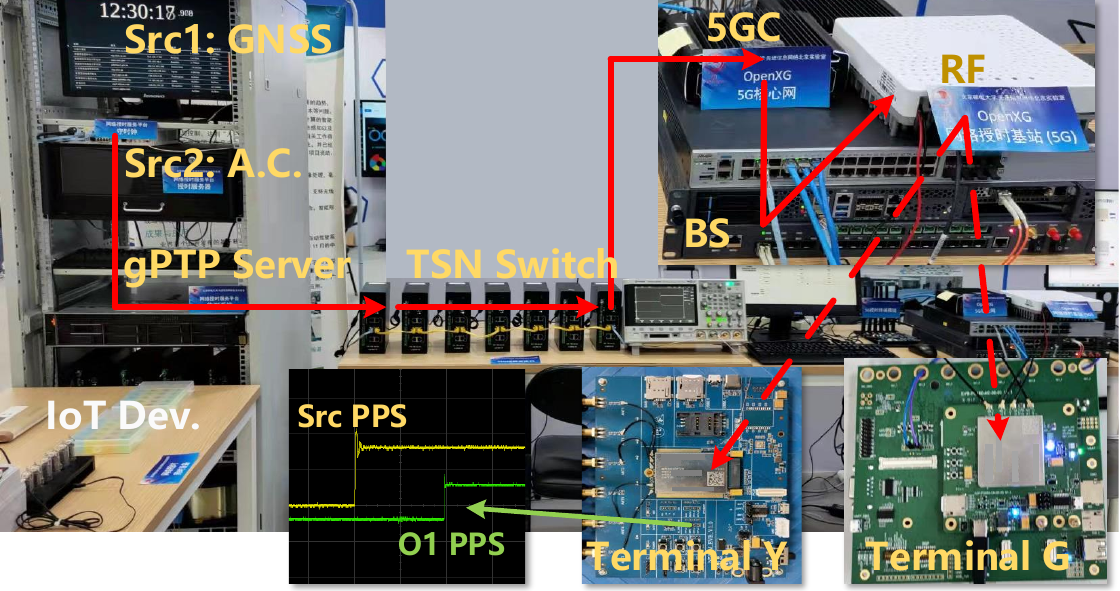}
  \end{center}
  \caption{The TAP prototype system and the topology of evaluation.}
  \label{fig-testtopo}
\end{figure*}

\begin{figure}[t!]
  \begin{center}
      \includegraphics[width=3.5in]{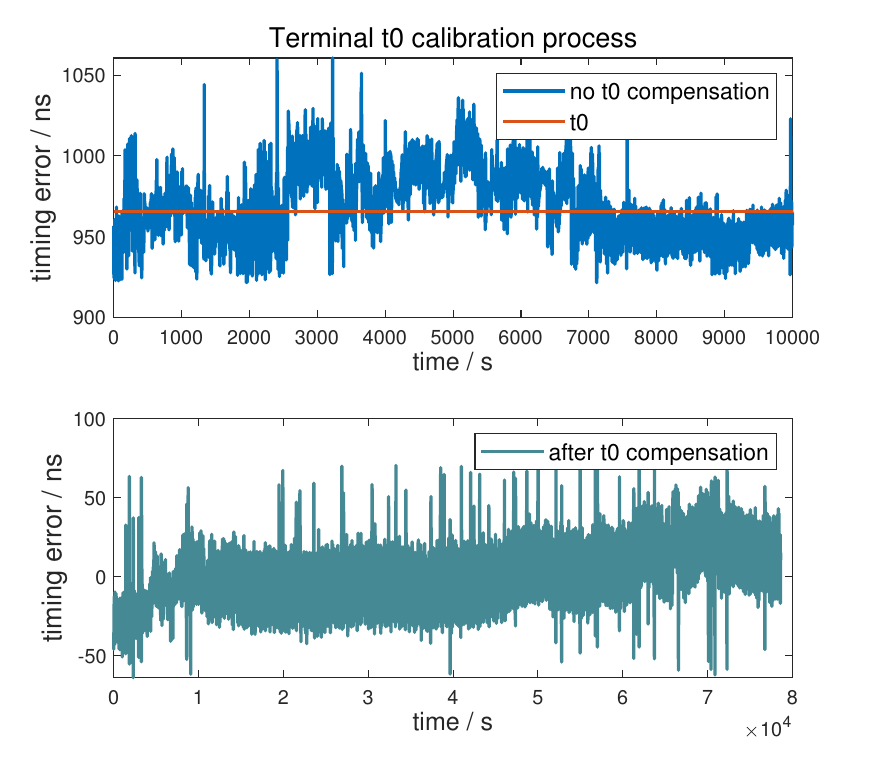}
  \end{center}
  \caption{The $t_0$ compensation of terminal $G$.}
  \label{fig-t0}
\end{figure}
\subsection{UE: Timing Module with Deterministic Latency}
When receiving timing info, UE not only needs to apply algorithms for wireless delay compensation but also needs to consider its own processing delay which includes baseband interaction delay, compensation algorithm delay and transcoding delay. Since this delay is difficult to predict, we propose to set a gate control timer with configurable value $t_0$ to compensate the processing delay of the module, which means that the transcode result $t$ is output as $t+1s$ after a delay of $1s-t_0$. $t_0$ needs to be obtained through calibration test.

The architecture of timing module is shown on the right in Fig. \ref{fig-bsue}. 

1) The baseband receives signals from air interface. The frame boundary is used to trigger the state machine of the timing module. Second, the timing module gets the received SIB9, current SFN, SIB9 offset, TA and additional compensation info from the baseband through AT command. Third, the security check is performed by comparing reference SFN with SFN to avoid retransmission and replay attacks and checking CRC embedded in SIB9 to avoid bit interference. 
2) Mechanism compensation and statistical compensation are performed as discussed in section II. Fig. \ref{fig-ue} shows the mechanism compensation workflow of the module. State $S_0$ is the initial state where any timing result will be accepted. State $S_1$ is the buffer status. When the subsequent timing result exceeds the threshold $Th0$, it will return to $S_0$, otherwise enters the locked state $S_2$ without accepting timing result. The buffer state can smooth the UE time adjustment and make the UE wait for the timing when the channel symmetry is relatively good. If any of the SFN check, CRC check and PRS-assisted error sensing fails, it interrupts the state machine to enter $S_1$. 
3) At last, the module outputs absolute time info according to the timer's control.

\section{Evaluation}
\subsection{Prototype System}
We implement a prototype system using software defined radio and some COTS hardwares. The grand master clock is based on rubidium clock OSA5421 which is synchronized to UTC. The BS protocol runs on Linux with low-latency kernel on x86 platform. Xilinx FPGA is applied for trigger and RF. The UE is based on U** V510 baseband, it integrates the timing module with IRIG-B transcoding module. We have developed two types of terminals, terminal $G$ only supports minimum timing function while terminal $Y$ supports full functions proposed in this paper. The protocol software, deployment guidelines, use case, etc. can be found at project \cite{opentap}, which is opensource under multiple licenses. 

\begin{table}[t!]
  \centering
  \caption{Parameter settings of the prototype system.}
  \label{table2}
  \begin{tabular}{|c|c|c|}
  \hline
  5GC                        & OpenXG                                                                               & /                                                                                \\ \hline
  \multirow{8}{*}{BS}        & Reference clock                                                                      & OSA5421-PTP                                                                      \\ \cline{2-3} 
                             & Platform                                                                             & \begin{tabular}[c]{@{}c@{}}Xeon Gold 6150, \\ Xilinx FPGA,\\ Linux\end{tabular} \\ \cline{2-3} 
                             & Band                                                                                 & n79, 30KHz SCS                                                                   \\ \cline{2-3} 
                             & SIB9 period                                                                          & $320ms$                                                                            \\ \cline{2-3} 
                             & $Gr$                                                                                 & $10ms$                                                                           \\ \cline{2-3} 
                             & $Sched\_pre$                                                                          & $20ms$                                                                           \\ \cline{2-3} 
                             & $t_c$                                                                                 & $4ns$                                                                            \\ \cline{2-3} 
                             & $t_p$                                                                                   & 1s-22ns                                                                          \\ \hline
  \multirow{7}{*}{UE}        & Baseband                                                                             & U*** V510                                                                        \\ \cline{2-3} 
                             & \begin{tabular}[c]{@{}c@{}}State \\ threshold\end{tabular}                    & \begin{tabular}[c]{@{}c@{}}Th0=2340ns(CP/2), \\  Th1=260ns(Gr)\end{tabular} \\ \cline{2-3} 
                             & \multirow{2}{*}{\begin{tabular}[c]{@{}c@{}}Statistical \\ compensation\end{tabular}} & G**: None                                                                          \\ \cline{3-3} 
                             &                                                                                      & Y*: Kalman                                                                        \\ \cline{2-3} 
                             & \multirow{2}{*}{$t_0$}                                                                  & G**: $965.4ns$                                                                        \\ \cline{3-3} 
                             &                                                                                      & Y*: $6716.5ns$                                                                        \\ \cline{2-3} 
                             & Transcode                                                                            & TTL IRIG-B                                                                       \\ \hline
  \multirow{2}{*}{Test Tool} & Oscilloscope                                                                         & DSOX3024T                                                                        \\ \cline{2-3} 
                             & Time Sync Analyzer                                                                               & XGTime SyncOne                                                                       \\ \hline
  \end{tabular}
  \end{table}

\begin{figure*}[t!]
  \centering
  \subfloat[The timing error of the prototype system. The UEs moved at random distance from the BS within 100m during the test.]{
  \includegraphics[width=0.48\textwidth]{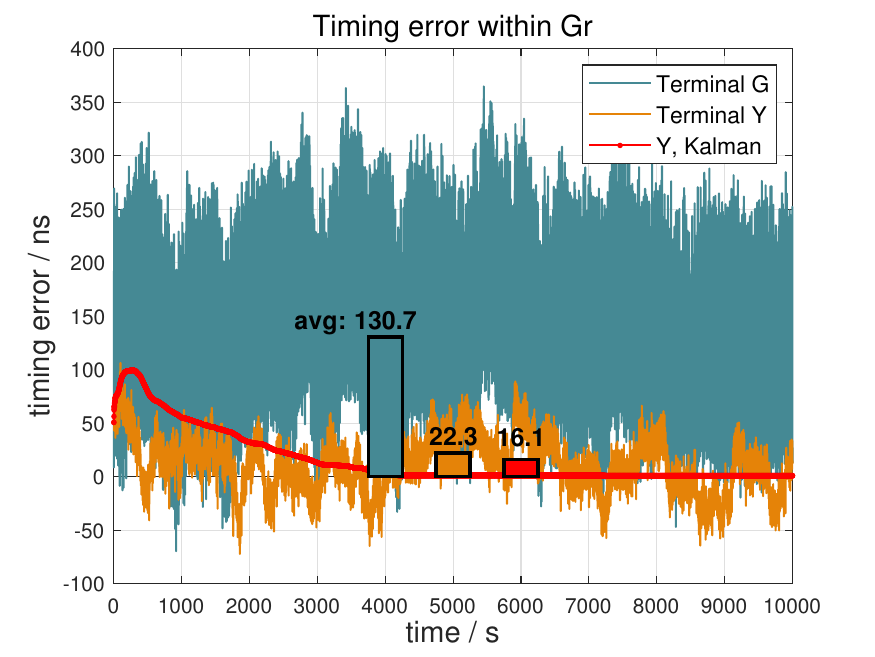}
  \label{a}
  }
  \hfill
  \subfloat[Comparison of timing error with and without statistical compensation of Terminal Y.]{
  \includegraphics[width=0.48\textwidth]{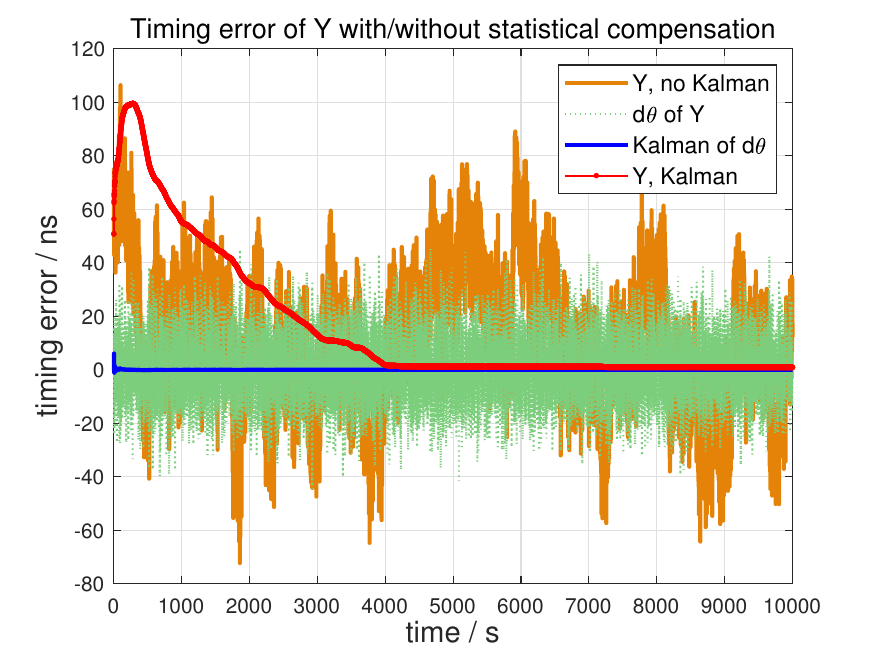}
  \label{b}
  }
  \quad
  \subfloat[Absolute timing error distribution of the prototype system compared with PTP.]{
  \includegraphics[width=0.48\textwidth]{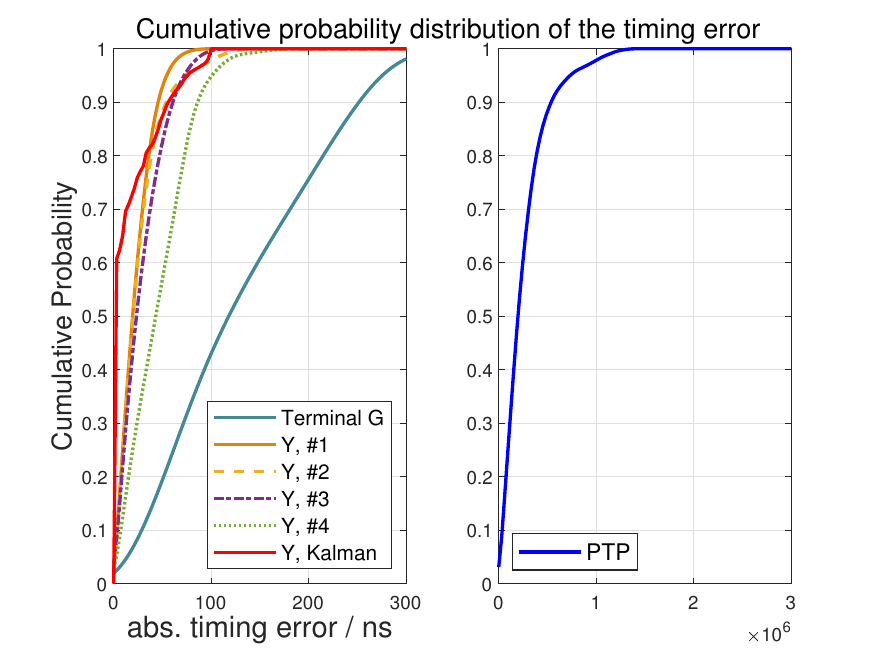}
  \label{c}
  }
  \hfill
  \subfloat[Allan variance of multiple observations of TAP-clock.]{
  \includegraphics[width=0.48\textwidth]{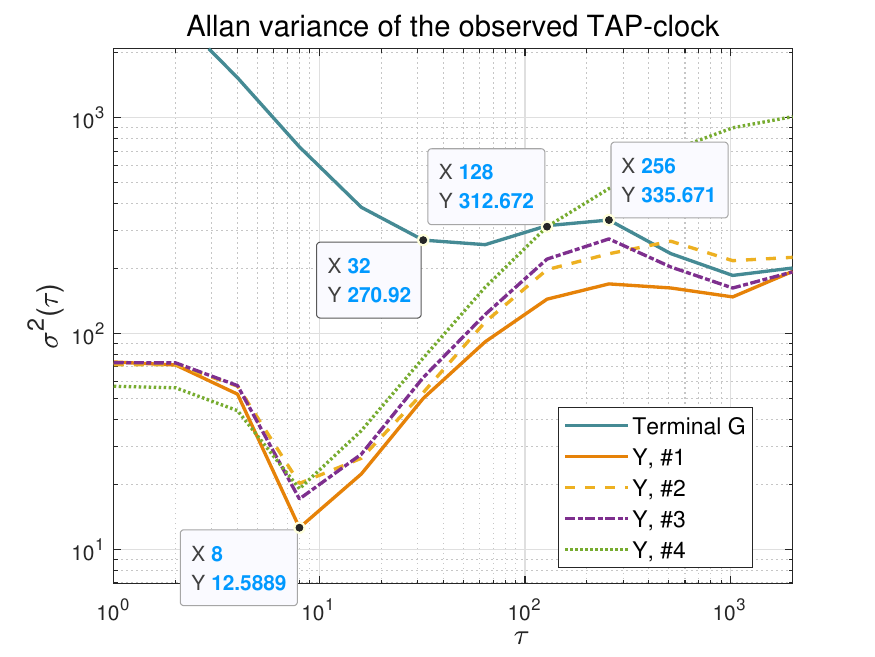}
  \label{d}
  }
  \hfill
  \subfloat[Boxchart of the timing error when using different value of $\tau$.]{
  \includegraphics[width=0.48\textwidth]{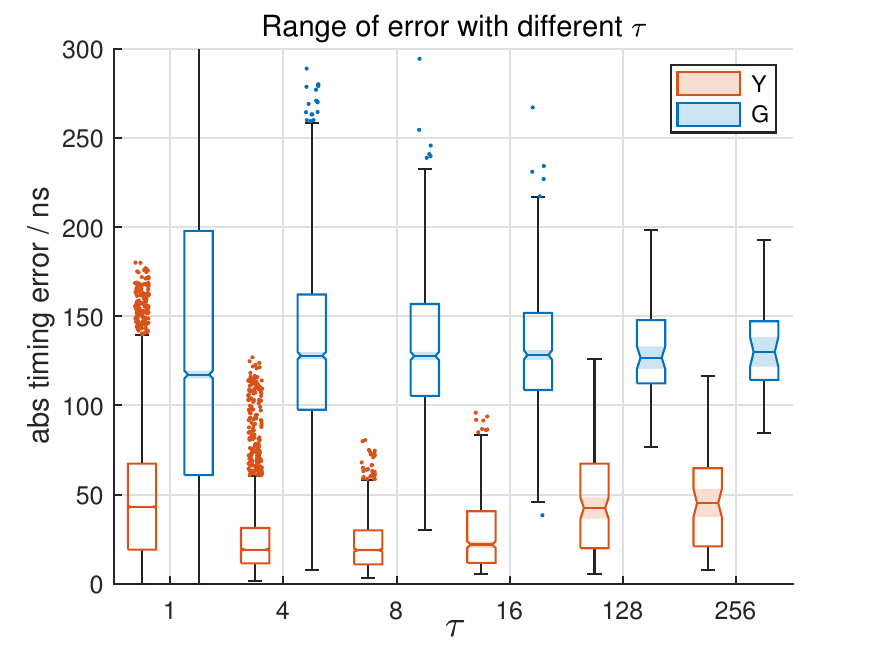}
  \label{e}
  }
  \hfill
  \subfloat[The remaining security problem of the proposed scheme.]{
  \includegraphics[width=0.48\textwidth]{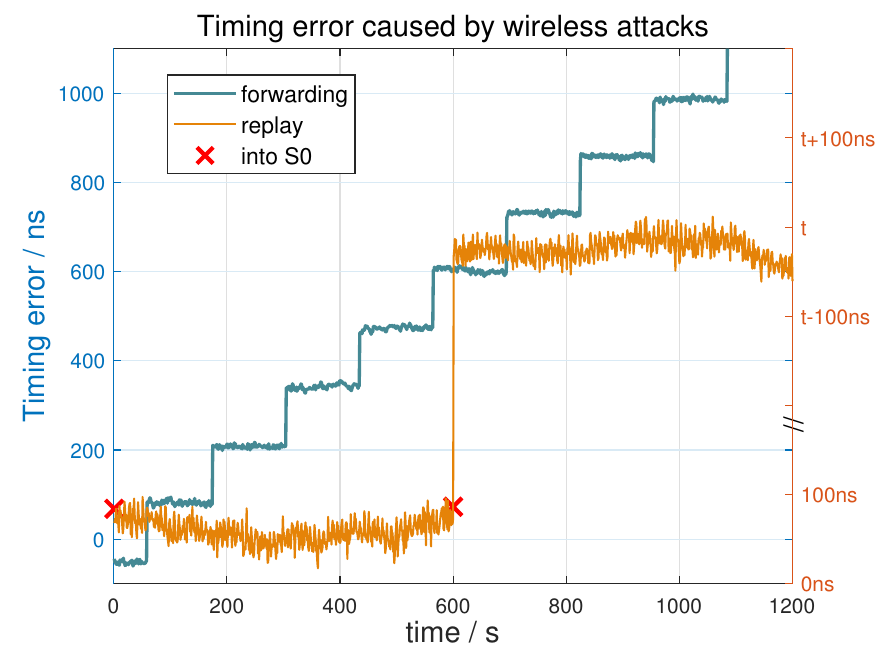}
  \label{f}
  }

  \caption{Test results of the prototype system.}
\end{figure*}

A typical TSN enabled IIoT scenario of the prototype is shown in Fig. \ref{fig-testtopo}. The compensation parameters for the system error including $t_c$, $t_p$, $t_0$ require to be determined by measurement before evaluation. 
\begin{enumerate}
  \item $t_c$: The trigger delay caused by the TTL rise delay can be obtained by viewing the PPS signal output by the FPGA through an oscilloscope.
  \item $t_p$: The FPGA PPS signal generating delay can be obtained by comparing the PPS output of master clock and FPGA.
  \item $t_0$: The protocol and calculating delay is also hardware-related. It can be obtained by connecting the BS and UE through and measuring the timing error without setting gate control timer. 
\end{enumerate}

Fig. \ref{fig-t0} shows the $t_0$ compensation process of terminal $G$. Without calibrating, the near-field timing error, $e$, ranges from about $900ns$ to $1100ns$, $t_0$ is calculated simply by: 
\begin{equation}
  t_0 = c|min[\sum abs(e-c)],c\in [min(e),max(e)]
\end{equation}
After configuring the gate control timer, the margin of error reaches about $\pm 150ns$ and the median is close to $0ns$. This completes the process of $t_0$ calibration.

Thereby, the settings of the prototype system are shown in Table \ref{table2}. 
We analyze the timing accuracy, stability and compatibility of the prototype system through E2E tests, and discuss the effectiveness of the TAP-clock model and TAP implementation method proposed in this paper. In the following, the ``timing error'' is obtained by comparing the PPS and ToD output of the master clock with the PPS and ToD encoded in IRIG-B output of the terminal.

\subsection{Timing Performance}
\subsubsection{Accuracy}
Fig. \ref{a} shows the basic timing error within $Gr$ of terminal $G$ and $Y$, where ``$+$" represents the terminal time lags behind the master clock and ``$-$" represents ahead. The timing error of $G$ is within the range of $-100$ to $+400ns$. The mean of its absolute error is $130.7ns$. The timing error of $Y$ without statistical compensation is in the range of $\pm 100ns$ with mean of $22.3ns$. The timing error of $Y$ with Kalman Filter is within the range of $+2$ to $+100ns$, it finally converges to $1.2ns$. Terminal $G$ does not apply SRS for delay compensation, thus when it moves in the cell, it cannot compensate for the downlink delay change with ns-level precision, causing the timing error to be biased towards ``$+$". Fig. \ref{a} does not show the $e_{Gr}$ of terminal $G$, i.e. several packet loss occurred and led to additional $10ms$ timing error. In contrast, $e_{Gr}$ of $Y$ is eliminated by the rSFN and security mechanism. The accuracy of $Y$ is limited by the channel delay estimation and the gPTP timing network between BS and MC. The statistical compensation allows $Y$ to converge to $10ns$-level accuracy after one hour of continuous observation, which is close to the long-term observation accuracy of GNSS. But still, it is contradictory to the short timing observation duration requirement of the general mobile terminals.

Fig. \ref{b} shows a more detailed Kalman-based compensating process. According to formula (\ref{f7}), we assume that $dT_{BS}(t)=0$ as atomic clock is applied (it is actually a value about $10^{-12}$. And if normal crystal oscillator is applied, it is likely to be 4$\sim$8 orders of magnitude higher than that of the atomic clock). The statistical compensation will track and filter $n_{\theta}[T_{BS}(t)]$ through $d\theta$ of terminal $Y$. In stead of accepting every timing result, $Y$ outputs according to the first timing result and accumulates the Kalman filter outputs. Note that it is sufficient to set the judgment condition for convergence to be no more than $\pm T_s/2$ within a certain period of time. Because error lower than that is random and indistinguishable, which is limited by the sampling rate of the mobile system. In this test, there are three multipaths between the terminal and the BS but no multipath hopping exists, the distribution of $d\theta$ is close to Gaussian distribution, thus the effect of Kalman is relatively good.

\subsubsection{Error distribution}
Fig. \ref{c} shows the absolute timing error distribution of $G$ and $Y$, where Y\#1$\sim$Y\#4 represents four times of test respectively, the signal-to-noise ratio of the terminal decreases successively. The test of transmitting PTP over the air interface serves as a comparison with the prior art. Without any optimization of PTP in mobile network, the accuracy of PTP is better than 1 microsecond with 97.8\% probability. It is difficult to meet some application requirements in IIoT that need $1\mu s$ synchronization accuracy. For the proposed prototype system, in the test with better SINR Y\#1, the absolute timing accuracy is better than $25ns$ with 50\% probability, better than $60ns$ with 90\% probability and 99.9\% better than $100ns$. While in the test with poor SINR Y\#4, the timing accuracy is better than $50ns$ with 50\% probability, better than $100ns$ with 90\% probability and 99.99\% better than $180ns$. Therefore, $Y$ can significantly achieve timing accuracy better than $\pm 200ns$. Due to the influence of the inability to accurately compensate the downlink delay, the timing accuracy of terminal $G$ can only be better than $150ns$ with 50\% probability, better than $250ns$ with 90\% probability and better than $400ns$ with 99.99\% probability. But still, it performs much better than the original-TAP and PTP. It is sufficient to meet the timing requirements of $1\mu s$.

\subsubsection{Allan variance}
Fig. \ref{d} shows the allan variance of $G$ and $Y$ according to formula (\ref{f8}). It can be seen that the allan variance of $Y$ reaches the minimum value around 20 when $\tau=8$. That is, the optimal timing stability can be achieved when accumulating $d\theta$ within 8 seconds (25 signaling) before outputting. The allan variance of $G$ rapidly decreases as $\tau$ increases to 32, and then it insignificantly changes until $\tau$ reaches 256. That is, the optimal observation time of $G$ is 4 times of $Y$ while its allan variance is over 20 times worse than $Y$. The technical characteristics of $G$ is similar to the original-TAP. The latter applies a parameter $K$ to control the duration of timing observation. When $K$ is set to 50, the timing stability can be significantly improved. And when $K$ is set to 125$\sim$175, the optimal average accuracy is obtained but the improvement is relatively less obvious. In our opinion,  $K$ is equivalent to the allan variance optimization, and the TAP-clock model proposed in this paper can effectively balance timing stability and observation duration.

Fig. \ref{e} shows the comparison of absolute timing error in multiple tests when applying different $\tau$. For terminal $G$, as $\tau$ increases, the median of timing error hovers around $125ns$. The interquartile range decreases thus the upper and lower quartile gradually move closer to the median. The maximum (worst) error decreases and the minimum error increases significantly. The number of outliers is small and can be basically ignored. The maximum error reaches $190ns$ at $\tau=256$. As for $Y$, there are a lot of outliers at $\tau=1, 4, 8$. And inferring from the notch, the median of $\tau=4, 8$ is lower than that of $1, 128$ and $256$ at a significant level of  95\%. The maximum error of $Y$ is lowest at $70ns$ when $\tau=8$ and increases up to $120ns$ when $\tau=256$. Apparently, excessive observation time is definitely not necessary for the terminals in motion as their channel environment is changing. Also, long observation time can be unacceptable for some specific application scenarios. In the timing of sub-microsecond magnitude, the occasional error of the delay estimation based on reference signal is relatively serious due to noise and multipath. We suppose that, to further improve the timing performance of mobile network, multiple antennas and improved pilot sequence are more important than other methods. 

\subsubsection{Security}
In addition to accuracy and stability of timing, security is also an important feature. Although the proposed scheme has improved security comparing with the original TAP, it still can be deceived in certain special cases. The blue line in Fig. \ref{f} shows a case of forwarding attack. A controllable reflection node receives the timing signaling and inserts a time offset of $130ns$ each time, which is half of the TAP control threshold $Th1$. If the main path of the current channel between the terminal and BS is blocked by some means, this reflection node can affect the $e_{inGr}$ of terminal $G$ or the SRS-based delay estimation of terminal $Y$, leading to the same result, that is, the timing error is gradually accumulated. The forwarding attack can lead to an error up to half the cycle prefix duration. Because when exceeding CP, the terminal will not be able to perform uplink synchronization and thus disconnect from the BS. It is difficult for the present standard single-BS timing scheme to deal with this situation, although the conditions for its implementation are harsh. The yellow line in Fig. \ref{f} shows a case of a continuous replay attack. Since the 5G system cyclically uses system frame numbers from 0 to 1023, if the timing signaling is replayed after one cycle, the terminal cannot judge it by the reference SFN in the signaling. Thereby a middleman, who is closer to the terminal and whose power is higher than that of the BS, continuously receives the timing signalings sent by the BS and continuously replays them after $10ms*1024=10.24$ seconds. Then it makes the terminal leaving the locked state $S2$ for initial state $S0$ and relocking to the middleman, resulting in an absolute timing error of $10.24s$. More efficient signaling identification is needed to defend against the replay attacks. These two attacking methods can obviously cause an error of more than $1\mu s$ and basically will not affect the normal communication process of the terminal. Their implementation difficulty is not significantly higher than that of GNSS signal interference. These problems still remain to be solved before TAP can be applied in the industries with high security requirements.
 
To sum up, the timing accuracy of the proposed prototype TAP timing system can fully meet the current 5G network synchronization accuracy needs of better than 1$\mu s$ with strong stability. Comparing the two terminal schemes of $G$ and $Y$, delay accurate estimation and SFN check have the most obvious improvement in timing stability and security. The former has lower hardware cost and is fully compatible with the R15 standard, the latter can improve several times of the timing performance.

\begin{figure}[t!]
  \begin{center}
      \includegraphics[width=3.5in]{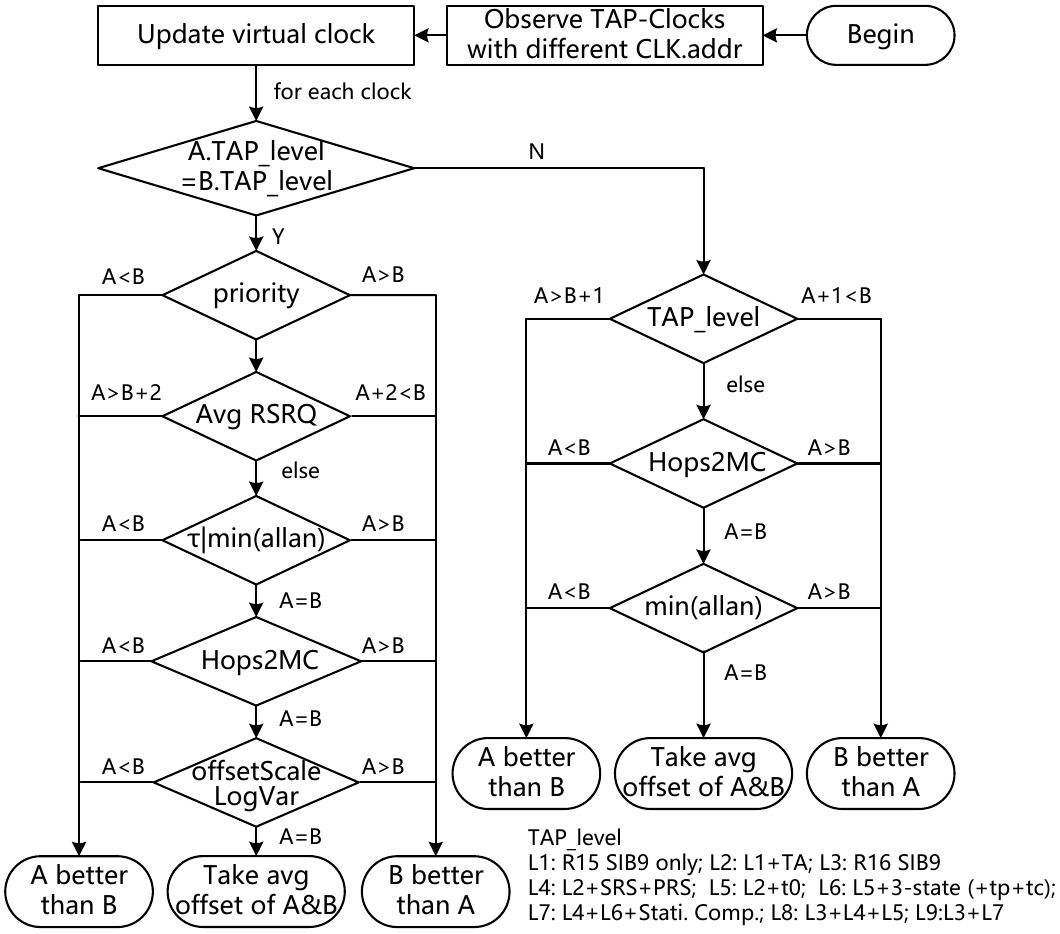}
  \end{center}
  \caption{Workflow of the Best TAP-clock Algorithm.}
  \label{fig-btca}
\end{figure}
\subsection{Compatibility and Coordination}
\subsubsection{5G-TSN Compatible} 
The convergence of 5G and TSN is one of the scenarios that requires the highest timing accuracy in the vertical applications. The TSN translator defined in TS 24.519\cite{i8} and TS 24.535\cite{i9} turns the 5G System into a TSN bridge. It makes the GMC transparent to the BS while the UEs are synchronized to GMC by gPTP over airinterface. To apply TAP to improve the timing accuracy of UEs, the BS can take the TSN GMC as a reference clock. That is, an additional TSN translator similar to DS-TT in \cite{i9} is required for the BS to communicate with GMC through gPTP. An offset of the BS time reference and GMC can be added to the SIB, thus BS does not need to synchronize to GMC as it is probably different from the reference clock of mobile network. For terminals, they can either obtain packets of GMC from the TSN translator and then replaces the time information by its local TAP-clock, or add offset to the gPTP packets. Few modification of the BS and UE is required for converging TAP and the present 5G-TSN. And if the 5G system itself is enabled by TSN, TAP is totally compatible. Our next step is to integrate TAP with 5G-TSN, part of the work has been carried out in the prototype system.

\subsubsection{Coordination between terminals}
Many IoT devices do not have mobile network communication capabilities. They are often accessed through mobile network gateways, which are usually called Customer Premise Equipments or Data Transfer Unit. Through the cooperation between these gateways, such as redundant transmission and backup access points, the reliability of the applications can be greatly enhanced. So does the absolute time synchronization. We propose a best TAP-clock algorithm which is similar to the Best Master Clock Algorithm (BMCA)\cite{bmca1}\cite{bmca2} of PTP. As shown in Fig. \ref{fig-btca}, each timing gateway in the same subnet sets its clock interface address as the following to avoid conflicts when using other network timing protocols:
\begin{equation}
  CLK.addr = CRC_{\{32-mask\}} (UE.id) + Addr.seg   
\end{equation}
where mask stands for the mask of the target address segment and UE.id can be the IP address in mobile network or UE RNTI. E.g., the UDP-carried PTP gateway address of UE with address $10.28.176.221$ to target IP segment $172.16.0.0/16$ is $172.16.0.0 + CRC_{32-16}(0A 1C B0 DD)=172.16.135.119$. Then, when multiple master clocks are observed by downstream devices, the clock source is chosen by applying the dataset comparison process. The first comparison factor is TAP\_level, which is determined by the enabled feature of TAP. Clocks with a level difference of no more than 2 need to continue to compare to the hops of the master clock Hops2MC, as each hop in the link introduces synchronization errors. If Hops2MC is the same, then a more stable (but not necessarily more accurate) clock can be selected by the allan variance. For clocks with the same level, signal quality and environmental changes affect the accuracy the most. Thus it compares the priority, average RSRQ, $\tau$ at the minimum allan variance, hops to MC and offset scale variance in order to select the best clock. If none of the factors can significantly distinguish one optimal clock, the average of all the best clocks is taken as a coordinated clock. 

\subsection{Future Work}
Since the timing based on TAP is accurate enough at present, stability and safety are the key points to ensure its application in the industrial field. Simultaneous reception of time signals from multiple clock sources and applying security checks is one of the necessary means. When a terminal encounters a single-point timing failure, how to obtain timing info of multiple cells and/or obtain time info from peers through Device-to-Device may be an effective method to alleviate its clock drift; When the terminal is moving, the compensation of the uplink and downlink delays is usually asymmetric. Improving the channel delay measurement during high-speed movement through multiple antennas and improving the design of the reference signal to reduce the Doppler effect on the delay measurement with low complexity is one of the effective methods to improve the applicability of TAP.

\section{Conclusion}
In this article, we have investigated the problems of high-precision mobile network absolute timing and discussed corresponding solutions based on TAP. Our contributions to the original TAP method include theoretically proving and improving the design of its mechanism and statistical compensation, which is validated for the first time by a prototype system. Specifically, first, we have analysed the end to end realization of TAP by establishing a TAP-clock model. We consider the relationship between the receiver's clock instability, air interface interference, system sampling ambiguity and the offset observed by the receiver each time, and give the expression of its Allan variance accordingly. Then we correlate the clock model with the mobile network error source. We have analyzed the clock source error, time information granularity error, air interface channel error and terminal error respectively, propose solutions to compensate for each error and give the upper and lower limits of the timing accuracy. Second, we have introduced an E2E implementation of TAP. It mainly includes schemes of high-precision time delivery and feedback for the BS and low-cost timing module with deterministic delay for the terminal. Corresponding to the clock model, we have discussed the hardware and software design from an engineering point of view. Third, we have developed a prototype system of the proposed schemes including a software defined radio based 5G BS and two models of COTS hardwares based 5G terminal. The results preliminarily show that the proposed scheme with full-function can achieve E2E timing error better than $\pm$$200ns$ with 99.99\% probability in various channel situations, while the scheme with minimum functional can achieve E2E timing error better than $-200\sim+400ns$ with 99.99\% probability. Through the analysis of allan variance, we concluded that under the condition of limited observation duration, there is an optimal solution for $\tau$ which can improve the timing stability under similar average timing error. The signaling cost is no more than $9.3kbps$ per cell or $106bps$ per UE while the additional hardware cost is almost negligible in mass production. Comparing with applying the existing timing methods in mobile network, it is significantly better than NTP, PTP and the original TAP in terms of accuracy and stability. It can break away from the reliance on GNSS with the help of technologies such as atomic clock and mobile edge computing. The major remaining issue is its security, the proposed scheme still can be affected by replay attacks and forwarding attacks under certain situations.

To conclude, the proposed methods can totally achieve sub-1-microsecond absolute time synchronization in 5G network, which is a basic requirement for time-sensitive vertical mobile applications in many industrial fields.


\section*{Note from 1st author}
This work was done at 2021.12, reviewed for about two years by TMC with some re-submission. Rejected at 2024.1 due to innovativeness issue. The long review procedure (with unprofessional review comments) made this work from the SoA (as we believed) to out-dated work with "just normal timing accuracy". 
Thus I decided to submit this work to arXiv. Although some features related to this work are already standardized by 3GPP in Release-17 (basically the PDC part), the prototype implemented in this work is still related to over 10 authorized patents. Please be careful when using the code of this work which was provided before with restricted license.
Any further action about this work, i.e., resubmission and modification, has nothing to do with the 1st author, who has already graduated.

\section*{Biographies}

\end{document}